\documentclass[%
 reprint,
showpacs,preprintnumbers,
 amsmath,amssymb,
 aps,
pra,
]{revtex4-1}
\usepackage{filecontents}
\usepackage{graphicx}
\usepackage{dcolumn}
\usepackage{bm}
\usepackage{hyperref}
\usepackage{natbib}

\begin{document}
\bibliographystyle{apsrev4-1}

\title{A mechanical switch for state transfer in dual cavity optomechanical systems}

\author{Satya Sainadh U}
\author{Andal Narayanan}%
 \email{andal@rri.res.in}
\affiliation{ Raman Research Institute, Bangalore, India.
}%


\date{\today}

\begin{abstract}
Dual cavity opto-electromechanical systems (OEMS) are those where two electromagnetic
cavities are connected by a common mechanical spring. These systems have been shown
to facilitate high fidelity quantum state transfer from one cavity to another. 
In this paper, we explicitly calculate the
effect on the fidelity of state transfer, when an additional spring is attached to 
only one of the cavities. 
Our quantitative estimates of loss of fidelity, highlight the sensitivity of dual cavity OEMS when it
couples to additional mechanical modes. We show that this sensitivity can be used to design an
effective mechanical switch, for inhibition or high fidelity transmission of quantum 
states between the cavities.
\end{abstract}

\maketitle


\section{\label{sec:level1} Introduction}
A corner stone optomechanical device is an optical cavity
attached with a mechanical spring to one of its cavity surfaces. The dynamical back action
resulting from coupling of the cavity mode with the mechanical mode
gives rise to a steady state, wherein, the resonance frequency of the cavity and the 
spring constant of the spring are altered ~\citep{kippenberg-science-321-1172}. 
Rapid progress in this field was made after the experimental demonstration of cooling
of the mechanical resonantor to its quantum mechanical ground state ~\citep{connell-nature-464-697}. 
The question of utility of such a device to store and transfer quantum states is an active
area of experimental and theoretical investigation, due to the ability of OEMS
to interface between different information processing modules. Indeed,
hybrid opto-electromechanical systems are fast becoming effective
lossless interfacing devices.\\
\indent A prototype opto-mechanical interface device was put forth by ~\citep{Tian,Regal,wang-prl-107-133601,clerk}, which
was subsequently experimentally realised ~\citep{oskar}. The effectiveness of this interface device
as a high fidelity quantum state transfer device ~\citep{wang-prl-107-133601}
is due to the existence of a dark state in the system Hamiltonian. This state 
does not include the mechanical
mode, thus minimising loss during state transfer. The dark state in such opto-mechanical
systems is quite analogous to the dark state present in atomic systems, which exhibit Electromagnetically
Induced Transparency (EIT) effect. Consequently, Optomechanically Induced Transparency (OMIT) effect
was predicted ~\citep{Agarwal} and observed ~\citep{Weis} in these systems. 
The search for quantum optics effects in 
these systems has given rise to theoretical predictions of wide variety of phenomena,
including slowing down of
a probe light ~\citep{tarhan} and Electromagnetically Induced Absorption (EIA)
~\citep{agarwal-pra-87-031802(R)} effect.\\ 
\indent In this paper, we focus on the consequences of coupling an additional
mechanical mode (spring 2 ) to one of the cavities of a dual cavity OEMS architecture (Figure 1). 
Such an analysis becomes highly relevent
when compact optomechanical sensors like a silicon microdisk  are being  engineered 
~\citep{karthik-opt-exp-20-18268} to interrogate another mechanical motion, like the motion of cantilever in
an Atomic Force Microscope. The mechanical motion of the cantilever couples to the optical modes
of the microdisk which can then be read out. However, the microdisk itself supports mechanical modes of its own,
whose frequencies can match that of the cantilever under study. In such systems, it is very pertinent to know
how the microdisk's mechanical modes couple through its optical modes, to the mechanical motion of the cantilever.
This situation maps to the architecture of a single optical cavity coupled to two springs which is a subset
of our dual spring dual cavity architecture (Figure 1).\\
\indent We show in this paper that, even one additional mechanical mode coupled to 
one of the cavities,
in a dual cavity OEMS,
 can reduce the fidelity of state transfer below 0.5. This feature allows use  
of the additional
mechanical mode as a switch, which either enables or inhibits 
high fidelity state transfer between cavities.
\section{\label{sec:level2} Model Hamiltonian}
\begin{figure}[h] 
\includegraphics[scale=.45]{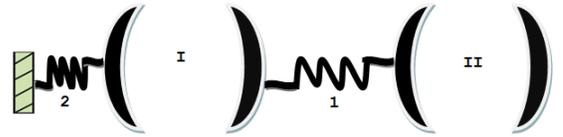}
\caption{Schematic of the dual spring-dual cavity optomechanical system. the first cavity is attached to two springs.}
\end{figure}
The dual cavity-dual spring system shown in Figure 1 is described by the Hamiltonian ($\hbar = 1$),
\footnotesize
\begin{equation}
\hat{H}=\sum_{i=1}^2\Big(w_m \hat{a}_i^\dagger \hat{a}_i-\Delta_i\hat{d}_i^\dagger \hat{d}_i+G_i(\hat{a}_1^\dagger \hat{d}_i 
+ \hat{a}_1 \hat{d}_i^\dagger )\Big)+G_3 (\hat{a}_2^\dagger \hat{d}_1 +\hat{a}_2 \hat{d}_1^\dagger)
\end{equation}
\normalsize
The cavities represented by the annihilation operators ($\hat{d}_i$),
can both be optical cavities or as is assumed in ~\citep{clerk-njp-14-105010},
 one optical and one microwave. 
The optomechanical coupling of both the cavities with the springs ($\hat{a}_i$) is provided by strong drive fields.
$w_m$ is the mechanical oscillation frequency of the springs 
  which is taken to be the same for both the springs. The drive fields are red tuned with $\Delta_i = -w_m$. 
The coupling constants are denoted by
$G_{i}$ = $c_{i} g_i$, where $c_i$s are proportional to the amplitude of the drive fields and $g_i$s 
denote the single photon coupling strength. 
$g_i$s are usually small, which results in the absence of quadratic cavity annihilation operator terms
in the cavity coupling. Thus, the coupling of the cavity modes  
with the mechanical modes are linear  ~\citep{prl-99-093901,prl-99-093902} in the Hamiltonian.
This Hamiltonian is written in a frame displaced by the strength ($|c_i|^2$) of the drive fields and 
in the interaction picture.
The internal losses of the cavities are denoted by
$\kappa_i$s and that of the springs by $\gamma_i$s. We take the good cavity limit with $G_{i} \gg \kappa_i$ and work in the 
resolved sideband regime with $\kappa_i \ll w_m$.\\
\indent In ~\citep{clerk-njp-14-105010}, the question of state transfer from cavity I to II was addressed
when both the cavities were coupled to a single spring. However, in realistic systems, there might be additional
 mechanical modes to which the cavity modes couple asymmetrically. 
In this paper, we address this question by  considering explicit coupling of the first cavity to
another spring (denoted by 2 in Figure 1). To simplify the calculation, we have made spring 2 to be identical to
spring 1.
\section{\label{sec:level3} Dynamical Evolution}
For explicitly studying the dynamical evolution of both cavity modes, we consider 
the hybrid scheme of ~\citep{clerk-njp-14-105010}, wherein the couplings $G_i$ are turned on simultaneously.
In the hybrid scheme, there is equal participation of dark and bright modes
during state transfer. In our architecture, due to the additional coupling, there exists no dark state for the
system Hamiltonian. In our calculations, we consider $G_1 = G_2 = G$ with
$G_3 = p G$ where $p$ is a tunable parameter.\\
\indent The absence of dark modes in the system's eigenstates, results in the mixing of cavity modes and mechanical modes
 during dynamical evolution of the system. Thus 
the states of the cavity modes cannot be swapped with each other due to contribution from mechanical modes. To explicitly see this, 
we evolve the cavity modes and mechanical modes using Heisenberg equations of motion, in the absence
of cavity dissipation $\kappa_i$ and mechanical dissipation $\gamma_i$. These are given by
\footnotesize
\begin{subequations}
\begin{widetext}
\begin{eqnarray}
e^{iw_mt}\hat{d}_2(t)=&&
\hat{d}_2(0)\left(\dfrac{\nu_- \cos\left(\dfrac{h_+t}{4}\right)+\nu_+\cos\left(\dfrac{h_-t}{4}\right)}{2}\right)+
\hat{a}_2(0)\dfrac{4ipG}{\sqrt{4+p^4}}\left( \frac{\sin\left( \frac{h_-t}{4}\right)}{h_-}-\frac{\sin\left( \frac{h_+t}{4}\right)}{h_+} \right)+
\hat{d}_1(0)\left(\frac{\cos\left( \dfrac{h_+t}{4}\right)-\cos\left( \dfrac{h_-t}{4}\right)}{\sqrt{4+p^4}} \right)\nonumber\\&&-
\hat{a}_1(0)(2iG)\left( \left( \nu_-+\frac{2}{\sqrt{4+p^4}}\right)\frac{\sin\left( \frac{h_+t}{4}\right)}{h_+}+ \left(\nu_+-\frac{2}{\sqrt{4+p^4}}\right)\frac{\sin\left( \frac{h_-t}{4}\right)}{h_-}\right),\\\nonumber\\
e^{iw_mt}\hat{d}_1(t)=&&
\hat{d}_1(0)\left(\dfrac{\nu_- \cos\left(\dfrac{h_-t}{4}\right)+\nu_+\cos\left(\dfrac{h_+t}{4}\right)}{2}\right)+
\hat{a}_1(0)\dfrac{i}{4G\sqrt{4+p^4}}\left(h_-\sin\left( \frac{h_-t}{4}\right)-h_+\sin\left( \frac{h_+t}{4}\right) \right)\nonumber\\+&&
\hat{d}_2(0)\left( \dfrac{\cos\left( \frac{h_+t}{4}\right)-\cos\left( \dfrac{h_-t}{4}\right)}{\sqrt{4+p^4}}\right)-
\hat{a}_2(0)(2iGp)\left( \nu_-\frac{\sin\left( \frac{h_-t}{4}\right)}{h_-}+\nu_+\frac{\sin\left( \frac{h_+t}{4}\right)}{h_+} \right),
\end{eqnarray}
\end{widetext}
\end{subequations}
\normalsize
where $\nu_\pm=1\pm\frac{p^2}{\sqrt{4+p^4}}$, 
$h_\pm=\sqrt{8G^2(2+p^2\pm\sqrt{4+p^4})}$. \\ 
We see from equation (2a) that at all times, there is non-vanishing contribution of
$\hat{a}_i$s to the state of second cavity $\hat{d}_2$. To address the question of state 
transfer we choose an optimum time given by $t_0=4\pi/h_+$.
For values of $p \gg 1$,
we have $\hat{d}_2(t_0)\approx\hat{d}_2(0)$ and $\hat{d}_1(t_0)\approx\hat{d}_1(0)$.
So it is clear that when the additional mechanical mode is strongly coupled to one of the cavity modes,
the state transfer gets totally inhibited. More importantly, we also see that the cavities retain the
initial state in which they were prepared. 
For small values of $p$, and for the particular case of $p=0$, at appropriate 
time $t_0$, we find 
$\hat{d}_1(t_0)=-\hat{d}_2(0)$ and  $\hat{d}_2(t_0)=-\hat{d}_1(0)$ (neglecting phase factors), thus 
recovering the results of hybrid scheme. \\
\indent We exploit this sensitivity of fidelity to additonal mechanical modes in a dual cavity OEMS, to outline 
a design for a mechanically mediated switch. This switch will facilitate high fidelity state transfer or totally inhibit it.
For practical implementation of this effect, we need an additional spring 2, whose spring constant 
can be varied externally. This can be done, for example, through application of an external voltage.
Initially, spring 2 is kept floppy so that no effective opto-mechanical coupling is established.
Thus state transfer between cavities proceed with high fidelity through spring 1.
Then by application of an external voltage the spring acquires a voltage dependent stiffness which
establishes effective opto-mechanical coupling. This, as we rigourously show below, reduces the fidelity of state 
transfer. We understand that electrostatic spring softening and 
stiffening structures, are already available in the field of MEMS ~\citep{mems-cite}, thus making practical
implementation of this idea feasible.\\
\indent Alternately, the coupling of spring 2 to cavity I can also be modified through its single
photon optomechanical coupling parameter $g_0$ where
$g_0 = x_{ZPF} \frac{\partial \omega_{cavity}}{\partial x}$ with $x_{ZPF} = \sqrt{\frac{\hbar}{2 m \omega_m}}$.
As is shown in ~\citep{karthik-njp-14-075015}, it is possible to tune $g_0$ of a cavity-spring
system through two orders of magnitude. 
So if we are to use this architecture for fabricating our dual cavity dual spring system, 
then we can achieve a large tuning range for $p$.

\indent In the following section, we show detailed calculations of fidelity, for intra-cavity state
transfer, for input Gaussian states in cavity I, in our dual spring, dual cavity architecture (Figure 1).
\section{\label{sec:level4} Fidelity calculation for input Gaussian states}
In this section, we give the details of our calculations and present the results for transfer fidelity,
for input Gaussian states. The input states are represented using Wigner functions and their dynamical
evolution is calculated using the Lindblad model for dissipation.
The cavity and spring systems are assumed to be in a bath at temperature $T$.
 The bath modes couple to the system modes giving rise to the density matrix $\hat{R}$
 which evolves according the equation $\dot{\hat{R}}=-i[\hat{H},\hat{R}]$.
 The reduced density matrix $\hat{\rho}$ , corresponding to the  resonator and cavity modes is obtained upon tracing out
 the bath degrees of freedom. $\hat{\rho}$ is expressed using the quadrature modes of cavity ($c$) and resonator $(m)$
 as $\bm{\hat{X}}^T \equiv (\hat{x}_{c_{k}},\hat{p}_{c_{k}},...,\hat{x}_{m_{k}},\hat{p}_{m_{k}})$, 
with $\hat{x}_{c_{k}}=\dfrac{\hat{d}_k+\hat{d}_k^\dagger}{2}$, $\hat{x}_{m_{k}}=\dfrac{\hat{a}_k+\hat{a}_k^\dagger}{2}$ and 
$\hat{p}_{c_{k}}=-i\dfrac{\hat{d}_k-\hat{d}_k^\dagger}{2}$, $\hat{p}_{m_{k}}=-i\dfrac{\hat{a}_k+\hat{a}_k^\dagger}{2}$, 
where $k$ can be 1/I or 2/II . The effective master equation for the reduced density matrix corresponding 
to the bilinear Hamiltonian form is given by,
\footnotesize
\begin{equation}
\dot{\hat{\rho}}=-i[\boldsymbol{\hat{X}}^T\bm{\hat{H}}\bm{\hat{X}},\hat{\rho}]+\sum_{j}\left(\dfrac{\Gamma_{j_+}}{2}\mathcal{D}\big( \bm{\hat{L}}^T_j\bm{\hat{X}}\big) + \dfrac{\Gamma_{j_-}}{2}\mathcal{D}\big( \bm{\hat{L}}^T_j\bm{\hat{X}}\big)^\dagger \right)\hat{\rho}.
\end{equation}
\normalsize
$\mathcal{D}\big( \hat{O}\big)\hat{\rho}\equiv 2\hat{O}\hat{\rho}\hat{O}^\dagger- \hat{O}^\dagger\hat{O}\hat{\rho}-\hat{\rho}\hat{O}^\dagger\hat{O}$
is the Lindblad superoperator. The $\bm{\hat{L}}^T_j\bm{\hat{X}} $ correspond to the mode annihilation operators of both the cavities
and the springs. $\Gamma_{j_\mp}$ denotes the rate of loss or amplification of energy from the bath into the system's $j^{th}$ mode, 
where $j$ can either be $c_{\mathrm{I}/\mathrm{II}}$($m_{1/2}$) corresponding to the first or second cavity (mechanical resonator). 
We consider the symmetric cavity case of $\kappa_1 = \kappa_2 = \kappa$. We also consider similar decay parameters for
both the springs with $\gamma_1 = \gamma_2 = \gamma$. For simplicity,
we consider the average number of thermal quanta exchanged by both the cavities with the bath to be the
same. The same holds true even for the springs; {\it i.e.} $N_{c_\mathrm{I}}=N_{c_\mathrm{II}}=N_c$ and $N_{m_1}=N_{m_{2}}=N_m$. 
With these parameters, we have $\Gamma_{c_{k}-}=N_c \kappa$, $\Gamma_{m_{k}-}=N_m \gamma$, $\Gamma_{c_{k}+}=(N_c+1)\kappa$ 
and $\Gamma_{m_{k}+}=(N_m+1)\gamma $. \\
We express the initial single mode Gaussian state in cavity $\mathrm{I}$ 
as a Wigner function $W_i(\bm{X})$. The Wigner function is characterized by the first moment of the mode quadratures 
$\overline{\bm{X}}_i$  and their covariance matrix $\bm{\sigma}_i$. From equation (3),
 one can evolve  $\overline{\bm{X}}$ and $\bm{\sigma}$ as,
\begin{eqnarray}
\frac{d \overline{\bm{X}}}{dt}=\mathcal{Q}\overline{\bm{X}}, \hspace*{.5cm} \frac{d \bm{\sigma}}{dt}=\mathcal{Q}\bm{\sigma} +\bm{\sigma} \mathcal{Q}^T +\mathcal{N}
\end{eqnarray}
The matrices $\mathcal{Q}$ and $\mathcal{N}$  are given by 
\footnotesize\begin{equation*}
\mathcal{Q}=\begin{pmatrix}
-\kappa/2 & 0 & 0 & 0 & 0 & G & 0 & pG \\ 
0 & -\kappa/2 & 0 & 0 & -G & 0 & -pG & 0 \\ 
0 & 0 & -\kappa/2 & 0 & 0 & G & 0 & 0 \\ 
0 & 0 & 0 & -\kappa/2 & -G & 0 & 0 & 0 \\ 
0 & G & 0 & G & -\gamma/2 & 0 & 0 & 0 \\ 
-G & 0 & -G & 0 & 0 & -\gamma/2 & 0 & 0 \\ 
0 & pG & 0 & 0 & 0 & 0 & -\gamma/2 & 0 \\ 
-pG & 0 & 0 & 0 & 0 & 0 & 0 & -\gamma/2
\end{pmatrix}
\end{equation*}
\normalsize\begin{equation}
\mathcal{N}=\dfrac{1}{4}Diag\{\bm{\overline{\kappa}},\bm{\overline{\kappa}},\bm{\overline{\kappa}},\bm{\overline{\kappa}},\bm{\overline{\gamma}},\bm{\overline{\gamma}},\bm{\overline{\gamma}},\bm{\overline{\gamma}}\}
\end{equation}
is a diagonal matrix,
with  $\boldsymbol{\overline{\kappa}}=\kappa(2N_c+1)$ and $\boldsymbol{\overline{\gamma}}=\gamma(2N_m+1)$. 
The matrix $\mathcal{Q}$ is responsible for the evolution of 
the system under the Hamiltonian including the intrinsic damping terms, 
while the matrix $\mathcal{N}$ consists solely of the bath parameters that determines the effect of bath on the system.  
Evolving the initial state  $W_i(\bm{X})$, using equations (3)-(5), we arrive at the final state $W_f(\bm{X})$,
at time $t_0$, in cavity $\mathrm{II}$ . 
The fidelity between these initial ($i$) and final ($f$) single mode Gaussian states is given by:   
\begin{subequations}
\begin{eqnarray}
F&=&\dfrac{1}{1+\overline{n}_h} \exp\left(-\dfrac{\lambda^2}{1+\overline{n}_h}\right) \hspace*{1cm} \text{with,}\\
\overline{n}_h&=&2\sqrt{Det[\boldsymbol{\sigma_i}+\boldsymbol{\sigma_f]}}-1 \\
\lambda^2 &=&\Big(\boldsymbol{\overline{X}_i}-\boldsymbol{\overline{X}_f}\Big).\frac{\sqrt{Det[ \boldsymbol{\sigma_i}+\boldsymbol{\sigma_f]}}}{ \boldsymbol{\sigma_i}
+\boldsymbol{\sigma_f}}.\Big(\boldsymbol{\overline{X}_i}-\boldsymbol{\overline{X}_f}\Big)
\end{eqnarray}
\end{subequations}
For simplicity, we consider the initial state to be a squeezed state given by,
\begin{equation}
\vert \alpha,r \rangle =\hat{D}(\alpha)\hat{S}(r)\vert 0 \rangle
\end{equation} with \begin{equation}
\hat{S}(r)=\exp\left( \frac{r}{2}\hat{d}_1^2-\frac{r}{2}\hat{d}_1^{\dagger 2}\right) 
\end{equation} where $r$ is real and \begin{equation}
\hat{D}(\alpha)=\exp\left(\alpha \hat{d}_1^\dagger-\alpha^*\hat{d}_1\right)
\end{equation} with $\alpha=|\alpha|e^{i\phi}$. Therefore the covariance matrix and first moment of the quadratures are given by,
\begin{eqnarray}
\boldsymbol{\overline{X}_i}=\begin{pmatrix}
|\alpha|\cos\phi \\ |\alpha|\sin\phi
\end{pmatrix}, \qquad
\boldsymbol{\sigma_i}=\frac{1}{4}\begin{pmatrix}
        e^{-2r} &0\\0&e^{2r}
        \end{pmatrix}.
\end{eqnarray} 
Based on mathematical formulation outlined above, we calculate 
the first moment and covariance matrix
for the output state. These work out to be
\begin{widetext}
\begin{eqnarray}
\boldsymbol{\overline{X}_f}&=&\underbrace{e^{-\dfrac{\text{\footnotesize$\kappa +\gamma$\normalsize}}{4}t_0 }\left( \frac{1+A_-}{\sqrt{4+p^4}}\right)}_{C_1} \boldsymbol{\overline{X}_i}
\\\nonumber\\
\boldsymbol{\sigma_f}&=& \Bigg[ e^
{
-\dfrac{\text{\footnotesize$\kappa +\gamma$\normalsize}}{2}t_0 
}
  \Bigg\{
        \frac{1}{16} \Big(
                           \nu_- - \nu_+ A_-(t_0) 
                     \Big)^2 
           +\frac{2G^2}{h_-^2} \left(
                                      \nu_+ - \frac{2}{\sqrt{4+p^4}}\right) \sin^2\left(\dfrac{h_-t_0}{4} \right)                                                                          
                               \Bigg\}+\boldsymbol{\overline{\kappa}}\:\Big( 
                                            \nu_+ \mathcal{I}_{2_{-}} + \nu_- \mathcal{I}_{2_{+}}
                                         \Big)
                                  \nonumber\\&&
                    \underbrace{\qquad\qquad\qquad\qquad\qquad\qquad\qquad+ \dfrac{2\boldsymbol{\overline{\gamma}}}{(h_+h_-)^2}
                                  \left(
                     \beta \Big( \nu_- \mathcal{I}_{1_{+}} + \nu_+ \mathcal{I}_{1_{-}} \Big)
                             +  \frac{2G^2(\kappa-\gamma)^2}{\sqrt{4+p^4}}                                
                      \Big(\mathcal{I}_{1_{-}}-\mathcal{I}_{1_{+}}\Big)
                      \right)\Bigg]}_{C_2}
          \boldsymbol{I}\nonumber\\&& +
          \underbrace{e^{-\dfrac{\text{\footnotesize$\kappa +\gamma$\normalsize}}{2}t_0 }\left( \dfrac{1+A_-}{\sqrt{4+p^4}}\right)^2}_{C_3}
  \boldsymbol{\sigma_i}  \end{eqnarray}
\end{widetext}
where the expressions  
$\beta=G^2\Big( 16G^2p^2-(\kappa-\gamma)^2 \Big)$,\\  $A_{\pm}(t)=\cos\left( \frac{h_{\pm}t}{4}\right)-\dfrac{\kappa-\gamma}{h_{\pm}} \sin\left( \frac{h_{\pm}t}{4}\right)$, and the integrals\\ \footnotesize $\mathcal{I}_{1_{\pm}}=\int_0^{t_0}e^{-\dfrac{\text{\footnotesize$\kappa +\gamma$\normalsize}}{2}t}
 \sin^2\left( \frac{h_{\pm}t}{4}\right)dt \quad$, 
\footnotesize$\mathcal{I}_{2_{\pm}}=\int_0^{t_0}e^{-\dfrac{\text{\footnotesize$\kappa +\gamma$\normalsize}}{2}t} A_{\pm}^2dt$
\normalsize 
\\
Identifying the coefficients of $\boldsymbol{\overline{X}_i}$, identity matrix ($\boldsymbol{I}$) and $\boldsymbol{\sigma_i}$
 as $C_1,C_2$ and $C_3$ respectively, we have:\\
\begin{equation}
\overline{n}_h=\frac{1}{2} \sqrt{(1+C_3)^2+16C_2^2+8C_2(1+C_3)\cosh(2r)}-1\end{equation} \footnotesize\begin{equation}
\lambda^2= \dfrac{2(1-C_1)^2|\alpha|^2}{\overline{n}_h+1}\Big( C_2+\dfrac{(1+C_3)}{4}(e^{2r}\cos^2\phi +e^{-2r}\sin^2\phi) \Big)
\end{equation}
\begin{widetext}
\begin{figure*}
 \includegraphics[scale = 0.35]{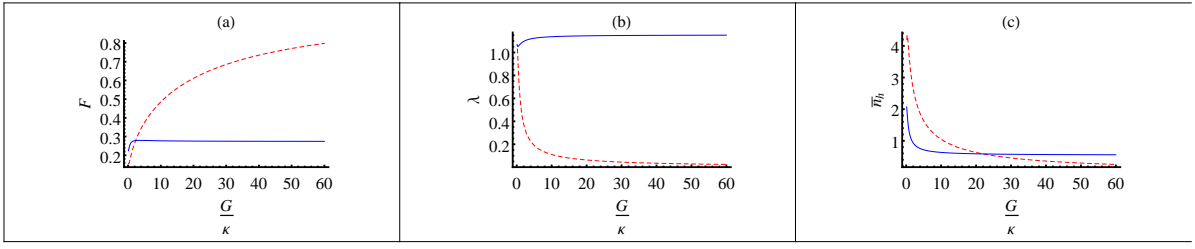}
\caption{Graphs 2(a)-2(c) show fidelity ($F$),amplitude decay ($\lambda$) and heat exchange with the bath ($\overline{n}_h$)
 respectively, 
 as a function of coupling strength ($\frac{G}{\kappa}$). This is drawn for an initial squeezed state
 of squeezing parameter $r = 1$, with $|\alpha|=1$,$\phi = \frac{\pi}{4}$ and with $p = 5$ (blue and thick). 
In all the three graphs a comparison plot is drawn for $p=0$ (red and dashed) which is the case where the 
extra spring 2 is absent. The graphs are plotted for experimentally realisable parameters which are,
 $\gamma=\frac{1}{50}\kappa=2\pi\times1KHz$, $\omega_m=2\pi \times10MHz$ with the symmetric cavity condition 
 \textit{i.e.} $N_\mathrm{I}=N_{\mathrm{II}}=N_c$, calculating for $\omega_{cavity}=2\pi \times10 GHz$ at $T_{bath}$ = 1.5 K.}
\end{figure*}
\end{widetext}
\normalsize
We see from the graph for fidelity (Figure 2(a)) that, with the additional spring, the fidelity falls to values
below 0.3 for strong couplings, over a wide range. The loss of fidelity 
is mainly due to the decay of amplitude denoted by the parameter $\lambda$ (Figure 2(b)), than from heat exchange with the bath
denoted by $\overline{n}_h$, (Figure 2(c)).
\begin{widetext}
\begin{figure*}
 \includegraphics[scale = 0.32]{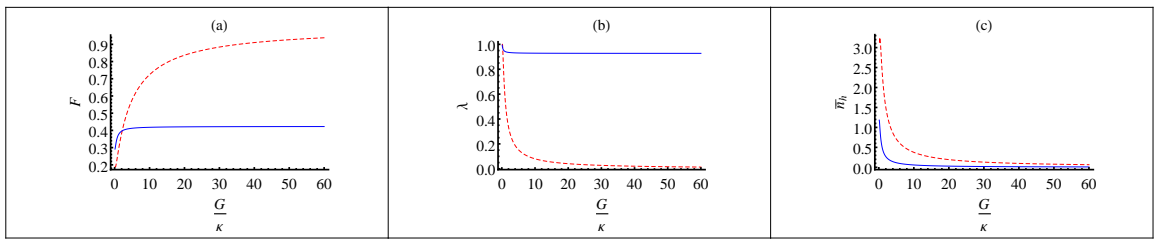}
\caption{The above graphs show fidelity ($F$),$\lambda$ and $\overline{n}_h$, as a function of coupling strength ($\frac{G}{\kappa}$),
 for a coherent state with $r=0$, $\alpha=1$ and for $p=5$ (blue and thick). Shown also in each graph are plots for $p=0$ 
 (red and dashed) which signifies the absence of the extra spring. The graphs are plotted for experimentally realisable parameters where,
 $\gamma=\frac{1}{50}\kappa=2\pi\times1KHz$, $\omega_m=2\pi \times10MHz$ with the symmetric cavity condition 
 \textit{i.e.} $N_\mathrm{I}=N_{\mathrm{II}}=N_c$, calculating for $\omega_{cavity}=2\pi \times10 GHz$ at $T_{bath}$ = 1.5 K.}
\end{figure*}
\end{widetext}
\normalsize
Figures  3(a)-3(c) show fidelity, amplitude decay and heat exchange, as a function of $\frac{G}{\kappa}$, 
for state transfer from cavity I to cavity II, for an input coherent state with $|\alpha| = 1$,for $p = 5$.
In comparison with corresponding graphs of Figure 2, we see that qualitatively, the coherent state
exhibits similar features to the squeezed state. Our calculations show that, for higher values of $p$, the fidelity for coherent states
reaches a limiting value of $(\frac{1}{e})^{|\alpha|^2}$, and is insensitive
to decay parameters of cavity and spring. In addition, for input coherent states, at very low bath temperatures around 0 K,
the heating parameter $\overline{n}_h$ $\rightarrow$ $0$ but for which
the amplitude decay ($\lambda$) is $\approx$ 0.9.
These features enable the action of the mechanical spring as 
a switch, to function with cavities of varying finesse and also at very low
ambient temperatures  ~\citep{stamper-kurn}.
\normalsize
\section{\label{sec:level5} Conclusions}
Dual cavity OEMS
are fast becoming model systems for high fidelity quantum state transfer.
In this report, we analyse and answer the very pertinent question of fidelity
of state transfer, when one of the cavities of this model system is coupled to an extra mechanical mode. We show that
the 
fidelity drops to a value below 0.5 for a wide range of values of coupling strength and decay
parameters. This highlights the need to isolate all 
spurious mechanical couplings in the design of interface optomechanical architectures.
Based on our calculations, we propose a mechanical switch which will  either
enable or inhibit high fidelity state transfer.  We envisage that the switch
can be made out of state of art MEMS actuators working on Electrostatic
Spring Softening mechanism.  
\bibliography{ref}
\end{document}